\documentclass[twocolumn,english,superscriptaddress,prl,amsmath,amssymb,floatfix,longbibliography]{revtex4-1}
\usepackage[T1]{fontenc}
\usepackage[utf8]{inputenc}
\usepackage{color}
\usepackage{babel}
\usepackage{amsmath}
\usepackage{graphicx}
\usepackage{esint}
\usepackage[unicode=true,pdfusetitle,
 bookmarks=true,bookmarksnumbered=false,bookmarksopen=false,
 breaklinks=false,pdfborder={0 0 0},pdfborderstyle={},backref=false,colorlinks=true]
 {hyperref}

\makeatletter

\usepackage{dcolumn}
\usepackage{bm}
\usepackage{color}
\usepackage{soul}
\usepackage{babel}
\usepackage{amsfonts}
\usepackage{slashed}
\usepackage{enumerate}

\usepackage{babel}

\makeatother

\begin{document}
\global\long\def\sgn{\mathrm{sgn}}%
\global\long\def\ket#1{\left|#1\right\rangle }%
\global\long\def\bra#1{\left\langle #1\right|}%
\global\long\def\sp#1#2{\langle#1|#2\rangle}%
\global\long\def\abs#1{\left|#1\right|}%
\global\long\def\avg#1{\langle#1\rangle}%

\title{Comment on ``How to observe and quantify quantum-discord states via
correlations''}
\author{Parveen Kumar \includegraphics[height=0.3cm]{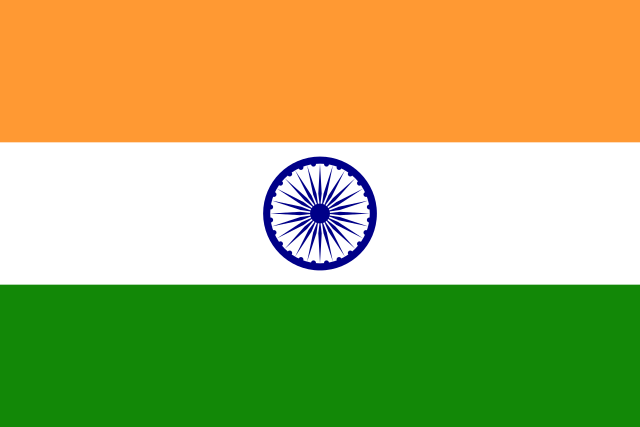}\,\,}
\affiliation{Department of Condensed Matter Physics, Weizmann Institute of Science,
Rehovot, 76100 Israel}
\author{Kyrylo Snizhko \includegraphics[height=0.3cm]{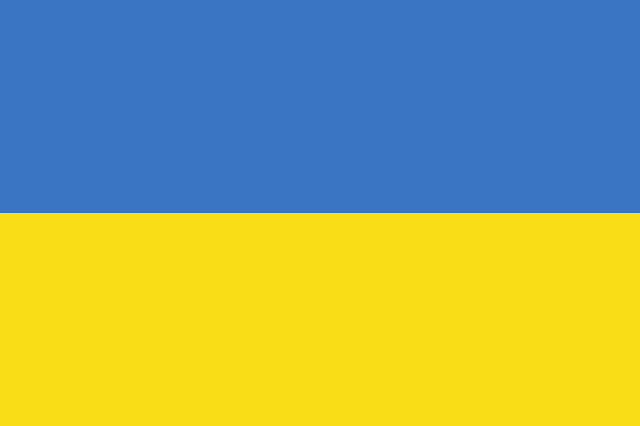}\,\,}
\affiliation{Department of Condensed Matter Physics, Weizmann Institute of Science,
Rehovot, 76100 Israel}
\begin{abstract}
The protocol proposed in the commented-upon article can certify zero
discord; however, it fails to \emph{quantify} non-zero discord for
some states. Further, the protocol is less efficient than calculating
discord via performing full quantum state tomography.
\end{abstract}
\maketitle
Recently Ref.~\citep{Hunt2019} proposed an interferometric protocol
for observing and quantifying quantum discord. The basic idea of the
protocol is that the shape of the zero visibility lines can distinguish
discorded states from the non-discorded ones, and (possibly) quantify
the amount of discord present in the state. Another advantage of the
protocol is that it maps the discord, a non-linear function of the
system density matrix, into direct results of measurements (which
are linear in the density matrix).

Two claims are made by Ref.~\citep{Hunt2019} regarding the proposed
protocol. (i) That the protocol enables one ``to detect and characterize
quantum discord of any unknown mixed state of a generic nonentangled
bipartite system'', namely to construct a ``discord quantifier {[}that{]}
is qualitatively consistent and quantitatively very close to the original
measure''. (ii) That the protocol is more efficient than calculating
the discord via full-state tomography of the system density matrix.
This statement follows, e.g., from the following passage in the introduction:
``Despite increasing evidence for the relevance of quantum discord,
\emph{quantifying} it in a given quantum state is a challenge. Even
full quantum state tomography would not suffice since determining
discord requires minimizing a conditional mutual entropy over a full
set of projective measurements. Moreover, even computing discord is
very difficult (it has been proven to be NP complete {[}18{]}). ...
In this paper, we propose an alternative discord quantifier which
would overcome these fundamental difficulties and render quantum discord
to be experiment friendly for many-body electronic systems.''

In this comment, we show that the protocol fails to be a \emph{universal}
discord quantifier. The comment is structured into three sections.
We first show, that the proposed protocol does indeed allow for idenitfying
non-discorded states. We then demonstrate that statement (i) is not
valid. Namely, we consider the example of the Werner state in which
the shape of the zero-visibility lines does not reflect the \emph{size}
of discord. Finally, we show that statement (ii) is not valid. Namely,
we demonstrate that the protocol is less efficient than calculating
the discord via full state tomography.

\emph{Section I. Identifying non-discorded states.---}The protocol
of Ref.~\citep{Hunt2019} is guaranteed to enable identification
of non-discorded states. This is stated in Ref.~\citep{Hunt2019}
and proven in the PhD thesis of M. Hunt \citep[Sections 5.3, 5.4]{Hunt2018}.
We find it useful, however, to briefly explain this fact here as it
elucidates the essence of the protocol discussed.

Consider a system comprising two parts, $A$ and $B$, in a state
characterized by density matrix $\rho^{AB}$. Similarly to Ref.~\citep{Hunt2019},
we assume that the Hilbert spaces of each $A$ and $B$ are two-dimensional.
It is known \citep{Ollivier2002} that the quantum state is not A-discorded
\emph{if and only if} the density matrix can be presented in the following
form:
\begin{equation}
\rho^{AB}=\sum_{i}p_{i}~|\psi_{i}^{A}\rangle\langle\psi_{i}^{A}|\otimes\rho_{i}^{B},\label{state_with_zero_discord}
\end{equation}
where states $\ket{\psi_{i}^{A}}$ form an orthonormal basis in the
Hilbert space of subsystem A, $\rho_{i}^{B}$ are arbitrary density
matrices of subsystem $B$, and $p_{i}\geq0$ are probabilities such
that $\sum_{i}p_{i}=1$.

Omitting unnecessary technicalities, the protocol of Ref.~\citep{Hunt2019}
involves the following elements: evolving $\rho^{AB}\rightarrow S\rho^{AB}S^{\dagger}$
with unitary $S=S^{A}\otimes S^{B}$, applying extra unitary $S_{d}\propto\exp\left(i\phi_{d}\sigma_{z}^{A}/2\right)$,
and measuring some observable involving subsystems $A$ and $B$ and
investigating its dependence on $\phi_{d}$, $S^{A}$, and $S^{B}$.
Assuming that $\rho^{AB}$ is of the form (\ref{state_with_zero_discord}),
one sees that there should exist $S_{0}^{A}$ such that $S_{0}^{A}\ket{\psi_{1}^{A}}=\ket{\uparrow}$
and $S_{0}^{A}\ket{\psi_{2}^{A}}=\ket{\downarrow}$. Since $S_{d}\ket{\uparrow}\bra{\uparrow}S_{d}^{\dagger}=\ket{\uparrow}\bra{\uparrow}$
and $S_{d}\ket{\downarrow}\bra{\downarrow}S_{d}^{\dagger}=\ket{\downarrow}\bra{\downarrow}$,
it is evident that $S_{d}S_{0}\rho^{AB}S_{0}^{\dagger}S_{d}^{\dagger}$,
where $S_{0}=S_{0}^{A}\otimes S^{B}$, does not depend on $\phi_{d}$
(i.e., any observable, e.g., $K_{\phi_{d}}$ from Eq.(5) of Ref.~\citep{Hunt2019},
will have zero visibility). Note that this does not depend on $S^{B}$
used. Therefore, in line with Ref.~\citep{Hunt2019}, \emph{if} the
state is non-discorded, the zero visibility lines in the protocol
do not depend on $S^{B}$. The correctness of the converse statement
is shown in the PhD thesis of M.~Hunt \citep[Sections 5.3, 5.4]{Hunt2018}.

\emph{Section II. Failing to quantify discord by zero visibility lines.}---The
above consideration shows that for non-discorded states the zero visibility
lines will not depend on manipulations with subsystem \textbf{$B$.}
In Section IV, Ref.~\citep{Hunt2019} suggests to quantify discord
(approximately) by looking at the deviation of the zero visibility
lines from horizontal (the $x$ axis is a parameter controlling $S^{B}$).
This suggestion is supported by a study of a small number of example
states. Here we provide an example of a family of states that have
different discord values yet all exhibit the same zero visibility
lines, which makes it completely impossible to \emph{quantify} the
discord even approximately using the quantifier proposed by Ref.~\citep{Hunt2019}.

Consider the Werner state
\begin{equation}
\rho_{W}(c)=c|\Psi^{-}\rangle\langle\Psi^{-}|+\frac{1-c}{4}\mathbb{I},\label{werner_state}
\end{equation}
where $|\Psi^{-}\rangle=\frac{1}{\sqrt{2}}\left(|\uparrow\downarrow\rangle-|\downarrow\uparrow\rangle\right)$
is the singlet Bell state, $\mathbb{I}$ is the identity matrix, and
$c\in[0,1]$. It is known \citep{Peres1996,Horodecki1996} that $\rho_{W}(c)$
is separable (or non-entangled) for $0\le c\le1/3$ and inseparable
(or entangled) for $1/3<c\le1$. In other words, for $c\leq1/3$,
the Werner state can be represented as a convex sum of product states
\citep{Wootters1998,Azuma2006}, cf.~Eq.~(1) of Ref.~\citep{Hunt2019}.
The states have non-zero discord for any $c>0$; the discord is given
by \citep{Luo2008,Ali2010}
\begin{equation}
\mathcal{Q}=\frac{1-c}{4}\log_{2}(1-c)-\frac{1+c}{2}\log_{2}(1+c)+\frac{1+3c}{4}\log_{2}(1+3c).\label{QD_werner_state}
\end{equation}
What is important for us is that the dependence of the discord on
$c$ is non-trivial.

\begin{figure}
\begin{centering}
\includegraphics[width=1\columnwidth]{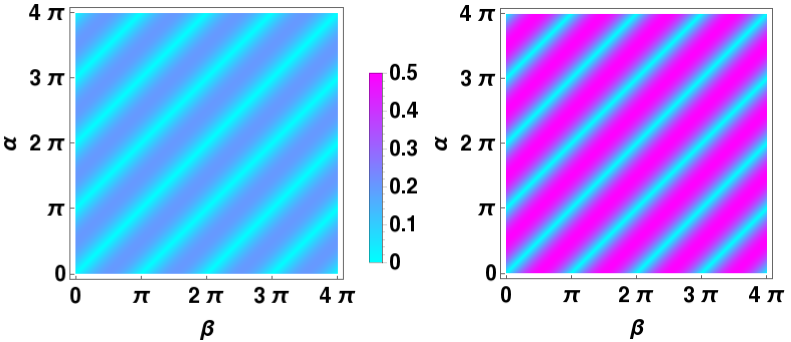}
\par\end{centering}
\caption{\label{fig:Werner_visibility_c_0p2.png}The dependence of the visibility
$\mathcal{V}(c)$, cf.~Eq.~(\ref{visibility_for_werner_state}),
for the Werner state on the parameters $\alpha$ and $\beta$ controlling
unitaries $S^{A}$ and $S^{B}$ (at $\phi_{A}=\phi_{B}$; $c=0.2$
(left) and $c=0.5$ (right)). The shape of the lines of zero visibility
does not depend on $c>0$.}
\end{figure}

The visibility in the protocol of Ref.~\citep{Hunt2019} applied
to the Werner state is \begin{widetext}
\begin{equation}
\mathcal{V}(c)=c\sqrt{(\sin(\alpha)\cos(\beta)-\cos(\alpha)\sin(\beta)\cos(\phi_{A}-\phi_{B}))^{2}+\sin^{2}(\beta)\sin^{2}(\phi_{A}-\phi_{B})},\label{visibility_for_werner_state}
\end{equation}
\end{widetext}where $\alpha,\phi_{A}$ and $\beta,\phi_{B}$ are
the parameters of $S^{A}$ and $S^{B}$ respectively, cf.~Eq.~(3)
in Ref.~\citep{Hunt2019} and the paragraph containing it. Observe
that the visibility depends on $c$ multiplicatively. Therefore, at
$c>0$, the shape of the lines of zero visibility does not depend
on $c$. In particular, for $\phi_{A}=\phi_{B}$, the zero visibility
lines in the $(\alpha,\beta)$ plane always correspond to $\alpha=\beta+\pi n$
with integer $n$, cf.~Fig.~\ref{fig:Werner_visibility_c_0p2.png}.
This shows that the shape of the zero visibility lines does not in
general allow for quantifying the discord even approximately.

\emph{Section III. The protocol efficiency.}---One of the motivations
mentioned in Ref.~\citep{Hunt2019} for proposing the discussed protocol
is finding an efficient way of estimating quantum discord. Here we
argue that the protocol of Ref.~\citep{Hunt2019} is computationally
less efficient than calculating the discord via full state tomography.
We start with the two-qubit case, considered in Ref.~\citep{Hunt2019},
and then provide estimates for the protocol complexity in the general
many-body context.

We first estimate the number of measurements required to obtain the
zero visibility lines via the protocol of Ref.~\citep{Hunt2019}.
Consider the case of the subsystems $A$ and $B$ having two-dimensional
Hilbert spaces. For each value of parameters $\alpha$, $\beta$,
$\phi_{A}$, $\phi_{B}$, $\phi_{d}$, a single measurement of the
observable yields a binary readout and the protocol utilizes the statistical
average of these readouts. In order to estimate the average to accuracy
$\epsilon\sim1/\sqrt{m}$, one needs to repeat the measurement $m$
times. Then, the same procedure has to be repeated for different values
of parameters. Suppose each parameter is sampled at $n$ points. This
yields $mn^{5}$ measurements that need to be performed when executing
the protocol.

We next estimate the cost of calculating the discord by means of full
quantum state tomography. Once the state density matrix $\rho^{AB}$
in a particular basis is known, quantum discord can be calculated
on a computer. One has to minimize the expression for discord (Eq.~(13)
of Ref.~\citep{Ollivier2002}) over the sets of projectors $\Pi_{i}^{A}=\ket{\psi_{i}^{A}}\bra{\psi_{i}^{A}}$
onto basis vectors of various bases in subsystem $A$ Hilbert space.
However, this operation can be performed using unitary rotations $S^{A}$
applied to $\rho^{AB}$ on a computer. Then it is enough to perform
15 different measurements on the system to recover $\rho^{AB}$. For
example, these can be measurements of three expectation values of
Pauli operators on subsystem $A$ ($\langle\sigma_{i=x,y,z}^{A}\rangle$),
three expectation values for subsystem $B$ ($\langle\sigma_{j=x,y,z}^{B}\rangle$),
and nine correlation functions $\langle\sigma_{i}^{A}\sigma_{j}^{B}\rangle$.
Assuming that one needs to estimate each of these averages to accuracy
$\epsilon\sim1/\sqrt{m}$, one needs to performs $15m$ measurements
to recover $\rho^{AB}$ and perform the calculation of its discord
on a computer.

Further, even if for some reason one does not want to perform the
basis optimization on a computer, one can do it via repeating the
tomography after applying various unitaries $S^{A}$, which is equivalent
to sampling parameters $\alpha$ and $\phi_{A}$. In this case, the
number of required measurement operations becomes $15mn^{2}$, which
is still less than $mn^{5}$ in the protocol of Ref.~\citep{Hunt2019}
for sufficiently large $n$. We thus conclude that estimating the
location of zero visibility lines in the protocol of Ref.~\citep{Hunt2019}
is less efficient than quantifying the discord by means of full quantum
state tomography.

Consider now the general case where the Hilbert spaces of subsystems
$A$ and $B$ have dimensions $d_{A}$ and $d_{B}$ respectively.
If they can be viewed as composed of $N_{A}$ and $N_{B}$ qubits,
then $d_{A}=2^{N_{A}}$ and likewise for $B$. Estimates similar to
the ones above show that the protocol of Ref.~\citep{Hunt2019} would
require $mn^{d_{A}^{2}+d_{B}^{2}-d_{A}-d_{B}+1}$ measurements, while
performing the full tomography including the minimization via explicit
sampling of $S^{A}$ would require $\left[(d_{A}+d_{B})^{2}-1\right]mn^{d_{A}^{2}-d_{A}}$
measurements. The last estimate does not count the operations with
the inferred density matrix on a computer; counting them increases
the last estimate by a factor which is only polynomial in $d_{A}+d_{B}$.
Therefore, for sufficiently large $n$, the full tomography is more
efficient than the protocol of Ref.~\citep{Hunt2019}. Similarly,
the protocol of Ref.~\citep{Hunt2019} exhibits worse scaling with
$d_{B}$. The only scenario in which the protocol of Ref.~\citep{Hunt2019}
may beat the full tomography is when keeping $d_{B}$ constant and
increasing $d_{A}$. However, even then the scaling is exponential
in $d_{A}$ (which is itself exponential in the subsystem size $N_{A}$)
and thus the protocol is not ``experiment friendly for many-body
electronic systems''.

\emph{Conclusion.} We have analyzed the protocol of Ref.~\citep{Hunt2019}
for quantifying quantum discord. We have shown, that despite it is
guaranteed to identify the states with zero discord correctly, it
cannot serve as a universal discord \emph{quantifier}. In particular,
on the Werner state, any quantifier based solely on the shape of zero
visibility lines in the protocol of Ref.~\citep{Hunt2019} fails
to capture the behavior of discord even qualitatively (apart from
saying whether it vanishes). Further, we showed that even if one disregards
this drawback, determining the shape of the zero-visibility lines
in the protocol of Ref.~\citep{Hunt2019} is less efficient than
performing full state tomography to calculate the discord.

At the same time we do recognize the physical value of the protocol.
While the discord is an abstract quantity, which is non-linear in
terms of the system density matrix, the protocol of Ref.~\citep{Hunt2019}
works only with directly observable quantities. Therefore, the protocol
of Ref.~\citep{Hunt2019} provides an intuitive illustration of what
it means for a state to have zero or non-zero discord.

\begin{acknowledgments}

We thank I.~Lerner, I.~Yurkevich, and Y.~Gefen for useful discussions.
We acknowledge funding by the Deutsche Forschungsgemeinschaft (DFG,
German Research Foundation) -- Projektnummer 277101999 -- TRR 183 (project
C01) and by the Israel Science Foundation (ISF).

\end{acknowledgments}

\bibliography{bibliography}

\end{document}